# Designing low-cost TaC virtual substrates for $Al_xGa_{1-x}N$ epitaxy

Dennice M. Roberts[1], Andrew Norman[1], Moira K. Miller[1,2], M. Brooks Tellekamp[1]

[1] National Renewable Energy Laboratory, Golden, Colorado 80401, USA
[2] Colorado School of Mines, Golden, Colorado 80401, USA

**Abstract**

$Al_xGa_{1-x}N$ is a critical ultra-wide bandgap material for optoelectronics, but the deposition of thick, high quality epitaxial layers has been hindered by a lack of lattice-matched substrates. Here we identify the (111) face of transition metal carbides as a suitable class of materials for substrates lattice matched to (0001) $Al_xGa_{1-x}N$ and demonstrate the growth of thin film TaC which has an effective hexagonal lattice constant matched to $Al_{0.45}Ga_{0.55}N$. We explore growth conditions for sputtered TaC on sapphire substrates and investigate the effects of sputter power, layer thickness and incident plasma angle on film structure and in- and out-of-plane strain. We then show critical improvements to film quality by annealing films in a face-to-face configuration at 1600 ºC, which significantly reduces full width at half max (FWHM) of in- and out-of-plane diffraction peaks and results in a step-and-terrace surface morphology. This work presents a path toward electrically conductive, lattice matched, thermally compatible substrates for $Al_xGa_{1-x}N$ heteroepitaxy, a critical step for vertical devices and other power electronics applications.

**Intro**

$Al_xGa_{1-x}N$ is a critical material for the increasing power electronics landscape due to its ultra-wide band gap (UWBG, 3.4 – 6.1 eV), bi-polar dopability, chemical and thermal stability, carrier mobility, and a mature thin film deposition technology. However, high quality $Al_xGa_{1-x}N$ has not yet reached its full potential due to a lack of suitable substrates necessary for maximum performance and scalable deployment. A lack of wafer-scale native substrates encompasses a large portion of the power electronics region of interest, as mapped in **Figure 1**.[1] Growth of $Al_xGa_{1-x}N$-devices by MBE and MOCVD is possible using non-native substrates, but dislocations from lattice-mismatch strain propagate into the $Al_xGa_{1-x}N$ layer and diminish device performance.[2–5] While nitride light-emitting diodes and some lateral devices are relatively agnostic to threading dislocations, vertical power electronics devices and deep-UV LEDs suffer from serious performance degradation from these defects which are generated in the mismatched heteroepitaxial process.[6,7] AlN and GaN are calculated to be thermodynamically fully miscible, but strain drives phase separation in this system particularly for thicker layers, limiting the practical composition range and further motivating the need for a lattice matched substrate.[8] Significant efforts in strain balancing and buffer layer engineering seek to reduce strain-related concerns and have successfully reduced dislocation defects to the low $10^8$ levels.[2,9,10] Nonetheless, these techniques significantly increase the complexity of device growth, do not capture the full range of useful $Al_xGa_{1-x}N$ compounds, and still limit device design and performance.[6,11] For example, crack-free $Al_xGa_{1-x}N$ layers have been achieved on sapphire substrates after patterning or on AlN substrates with precise and complex heterostructure growth, but these methods remain limited to finite layer thicknesses.[2,10,12] Critically, these buffer layers are insulating, preventing vertical device architectures which are necessary for high-current applications.

Suitable lattice-matched substrates require both nearly-identical lattice parameters and an appropriate thermal expansion coefficient to mitigate crack formation during growth or operation.

Work from Amano *et. al*. demonstrates successful epitaxial growth of GaN on hexagonal ZrB2 (0001).[13,14] This work illustrates the breadth of chemical space that could be explored for lattice-matched substrates. It also identifies another important factor: the chemical compatibility of the substrate and growth environment must be considered. In the case of $ZrB_2$, a low temperature GaN buffer layer was required to prevent nitridation of the boride surface from $NH_3$ at high temperatures.[15]

In contrast to bulk substrates, "virtual" substrates utilize thin, high-quality layers grown on inexpensive substrates to serve as a lattice-matched template for subsequent III-N growth. Miyake *et. al* shows growth of AlN virtual substrates on sapphire, that, after annealing, enables homoepitaxial AlN growth with threading dislocation densities (TDD) of $1.7 \times 10^8$ cm$^{-2}$.[16] While these virtual substrates are still deposited on sapphire, the dislocation annihilation occurs within 100's of nm rather than the ~10 μm dislocation-terminating AlN layer used in MOCVD-nucleated structures. The elimination of this thick buffer layer significantly reduces the threat of cracking due to thermal expansion mismatch.

Using the above cases as inspiration, we are exploring the (111) plane of the transition metal carbides and nitrides as a family of compounds suitable for substrates throughout the AlN-GaN composition range. As a case study we investigate growth of TaC which possess an ideal (111) lattice spacing matching the basal (0001) plane of $Al_{0.55}Ga_{0.45}N$. In addition to the in-plane lattice matching, the out-of-plane TaC step height (2.56 Å) is nearly-matched to the bilayer thickness of $Al_{0.55}Ga_{0.45}N$ (2.53 Å). **Figure 1** demonstrates the relationship between effective (111) lattice parameter of cubic TaC and hexagonal $Al_xGa_{1-x}N$. We note that the lattice parameter of TaC can also be modified by deposition conditions and stoichiometry, as the rock salt structure is capable of stably hosting up to 50% carbon vacancies.[17–19] This composition range is important for power electronics applications due to its large potential critical electric field and acceptable carrier mobility providing a large Baliga figure of merit (BFOM). The $Al_xGa_{1-x}N$ BFOM meets or exceeds that of GaN with increasing Al content beginning at 30 % - 60% Al depending on temperature and mobility calculation method.[20,21] In addition to an increased BFOM $Al_xGa_{1-x}N$ devices are theoretically capable of higher voltage operation than GaN (and comparable to or exceeding β-$Ga_2O_3$ with the added benefit of p-type doping), benefits which are currently unobtainable due to the aforementioned lack of suitable substrates.[21] The mismatch of coefficient of thermal expansion between TaC and $Al_xGa_{1-x}N$ between room temperature and $Al_xGa_{1-x}N$ growth temperatures (~1150-1350 ºC via MOCVD) is around 0.2% for most Al-compositions, indicating $Al_xGa_{1-x}N$ films grown on TaC are unlikely to crack during cooldown after epilayer growth.[19] Additionally, TaC is highly conductive (σ ≈ 4.5 ∗ 10$^6$ S/m at 20 °C)[22,23] and used in high temperature applications such as coatings and ohmic contacts in SiC electronics.[24–27] Electrically conductive substrates are a critical component of vertical power devices; a conductive substrate reduces complexity and processing steps needed to contact the bottom terminal of the device.[28] Substrates with metallic conductivity are more desirable than highly doped semiconductor substrates because the ~2-4 orders of magnitude decrease in resistivity significantly decreases the on-state resistance which is inversely proportional to the effective BFOM.

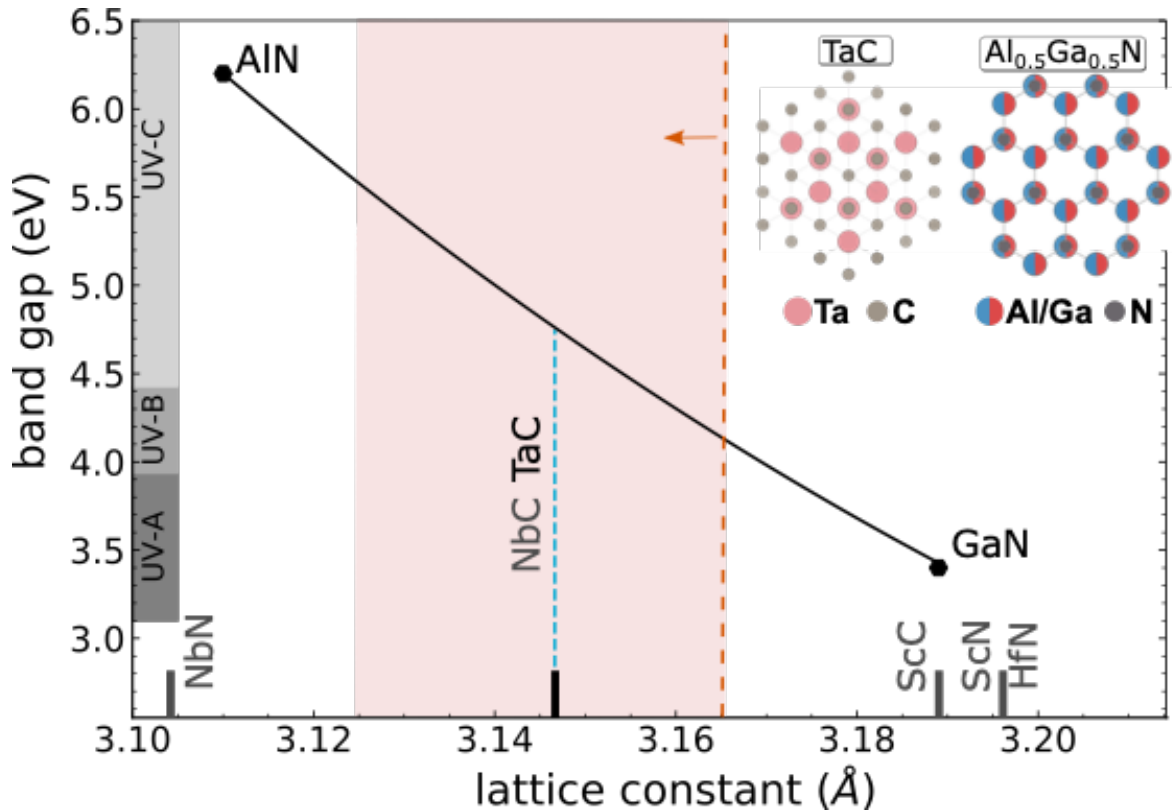

*Figure 1. Lattice constants of Al$_x$Ga$_{1-x}$N compounds compared to transition metal carbides and nitrides. The red shaded region and dotted line represent the region of the Al$_x$Ga$_{1-x}$N space particularly useful for power electronic applications, where the BFOM is predicted to exceed that of GaN in certain operating conditions. Structure of the TaC (111) and Al$_{0.5}$Ga$_{0.5}$N (0001) faces shown at upper right. The horizontal axis shows the in-plane lattice constant for (0001) oriented AlN-GaN phases and the effective in-plane lattice constant for (111) oriented rock salt carbides and nitrides.*

TaC can be grown by sputtering, evaporation, chemical vapor deposition (CVD), physical layer deposition (PLD), and ion-assisted deposition.[10–12] In this work we grow epitaxial (111)-oriented TaC on sapphire substrates by RF sputtering. We explore the effect of various sputter conditions on film quality and determine improved crystal quality and phase purity at low working pressures and high cathode powers. Global film stoichiometry is close to 1:1 and uniform throughout the thickness of the film as confirmed by Rutherford backscattering spectroscopy (RBS). Further, we show the TaC thin film structure depends on film thickness and plasma incidence angle during growth. Further, we explore changes to crystal structure after annealing in an inert environment at 1500 °C - 1600 °C. X-ray diffraction (XRD) patterns of annealed films show reduced full-width half-max (FWHM) for both in-plane and out-of-plane peaks, suggesting increased grain sizes and improved crystal quality. Surface morphology of films before and after annealing is assessed by atomic force microscopy (AFM), and the crystalline structure of the film and interface are imaged by high-resolution transmission electron microscopy (HR-TEM). This work illustrates the potential of TaC and other transition metal carbides for use as lattice-matched virtual substrates for Al$_x$Ga$_{1-x}$N devices.

**Results and discussion**

*Growth of TaC thin films |* We explore the effect of several sputter process conditions to tune crystallinity and suppress competing phases. Based on literature studies of TaC growth indicating higher temperatures stabilize the (111) face of TaC and our own initial studies, the substrate growth temperature is fixed at 730 °C.[27] This is the highest achievable temperature for the growth chamber at the time of these experiments, however we expect further improvements to film quality will be achievable with higher substrate temperatures. We choose a low working pressure of 5 mTorr as minimizing pressure can reduce stress in the growing film by enabling higher adatom mobility at the substrate surface and is favorable for maintaining oriented grains.[31,32] **Figure 2** shows the effect of sputter gun power on film quality and a detailed structural analysis of the crystal structure of the grown film. The effect of sputter cathode power on film structure is shown in **Figure 2(a)**. A range of 20W to 60W is explored; note that for the target used, power density requirements limit the maximum power to approximately 65 W.

**Fig 2(a)** shows XRD patterns of thin films deposited at 20 W have peaks shifted toward higher $2\theta$. This shift and the presence of a shoulder around 35° indicates the presence of a multi-phase system, likely the (002) plane of hexagonal $Ta_2C$. This is consistent with other reports of TaC thin films an "intermediate disordered phase" of other Ta-C compositions, primarily $Ta_2C$.[24,30] For reference, the lattice parameter of this phase is less desirable for $Al_xGa_{1-x}N$ as it is 0.3103 nm, smaller than that of binary AlN,[33] although conductive bottom-contacts for AlN are a consideration. Recently-published AlN with p- and n-type conductivity provide a potential use-case for conductive epitaxial templates lattice-matched to AlN (mismatch to $Ta_2C$ is 0.2%).[34] As cathode power increases to 30 W, the signal from the competing $Ta_2C$ phase reduces, with the (002) shoulder still visible as asymmetric broadening but with lower intensity. As power increases, the XRD pattern approaches that of (111) TaC. Between 30 W and 60 W, simultaneous with a thickness increase, peak position shifts to larger lattice constants more akin to bulk TaC and peak intensity increases concurrently. From these patterns we calculate a 0.08 Å shift toward larger lattice constants in the out-of-plane lattice spacing for sputter cathode powers between 30 W and 60 W. Unless otherwise noted, we use a deposition power of 40 W for the remaining samples described in this work due to the quality of the (111) peak and an in-plane lattice parameter closer to the high-Al content $Al_{1-x}Ga_xN$.[1] **Fig 2(b)** shows a phi scan about the (-102) plane of the sapphire substrate and the (002) plane of TaC. We observe a 30° in-plane rotation between substrate and film at these reflections indicating an alignment of TaC to the oxygen sublattice of $Al_2O_3$ as is observed for GaN and AlN when nucleated on $Al_2O_3$.[35] We also observe six-fold coordination, rather than three-fold, indicative of twinning in the TaC crystal where the alternating intensity with φ indicates one orientation is slightly preferred over the other. This can be understood by looking at a (0001) slice of the $Al_2O_3$ crystal (**Fig 2c**) where there are 6 semi-equivalent oxygen sites (they are displaced in the c-direction). The observed heteroepitaxial relationship is therefore $(111)_{TaC} \parallel (0001)_{Al2O3}$ and $\{0\bar{1}1\}_{TaC} \parallel (1\bar{1}00)_{Al2O3}$.

To understand crystal quality of films and further ensure the observed peaks are associated with TaC rather than a competing Ta-C phase, we measured symmetric and asymmetric rocking curves for the 40 W films. **Fig 2(c)** shows the crystallographic structure of cubic TaC and the locations of the (111) and (113) planes, shown in yellow and red respectively. The angle between these planes is 29.5° for a cubic crystal. This angle is not achievable by a combination of any two planes in the $Ta_2C$ structure, so the presence of both planes verifies the presence of cubic TaC phase (although it does not rule out the presence of $Ta_2C$). **Fig 2(d)** shows the (111) plane of TaC in the expected location in omega. The (111) peak is well fit by two components – a sharp peak with a FWHM of 510 arcsec and a broad peak with FWHM of 4330 arcsec. This can be attributed to a two-layer system caused by mismatch between the substrate and growing film wherein one layer is strained and the other relaxed. This behavior, common in sputtered films, is found in literature on a similar AlN/sapphire thin film system and is discussed further below.[36] **Fig 2(e)** shows the existence of a strong (113) peak in the expected location in reciprocal space and a FWHM of 4330 arcsec, which combined with the (111) rocking curve is indicative of columnar grains with low tilt which are twisted around the (111) axis[37]. The calculated lattice parameter assuming a cubic system is 4.44 Å; however, as discussed below, a rhombohedral system may more accurately reflect the in-plane distortion that results from in-plane strain. Altogether, XRD data confirm the presence of epitaxial TaC oriented in the (111) direction, consistent with columnar sputtered growth with rotated and tilted grains above a critical thickness.

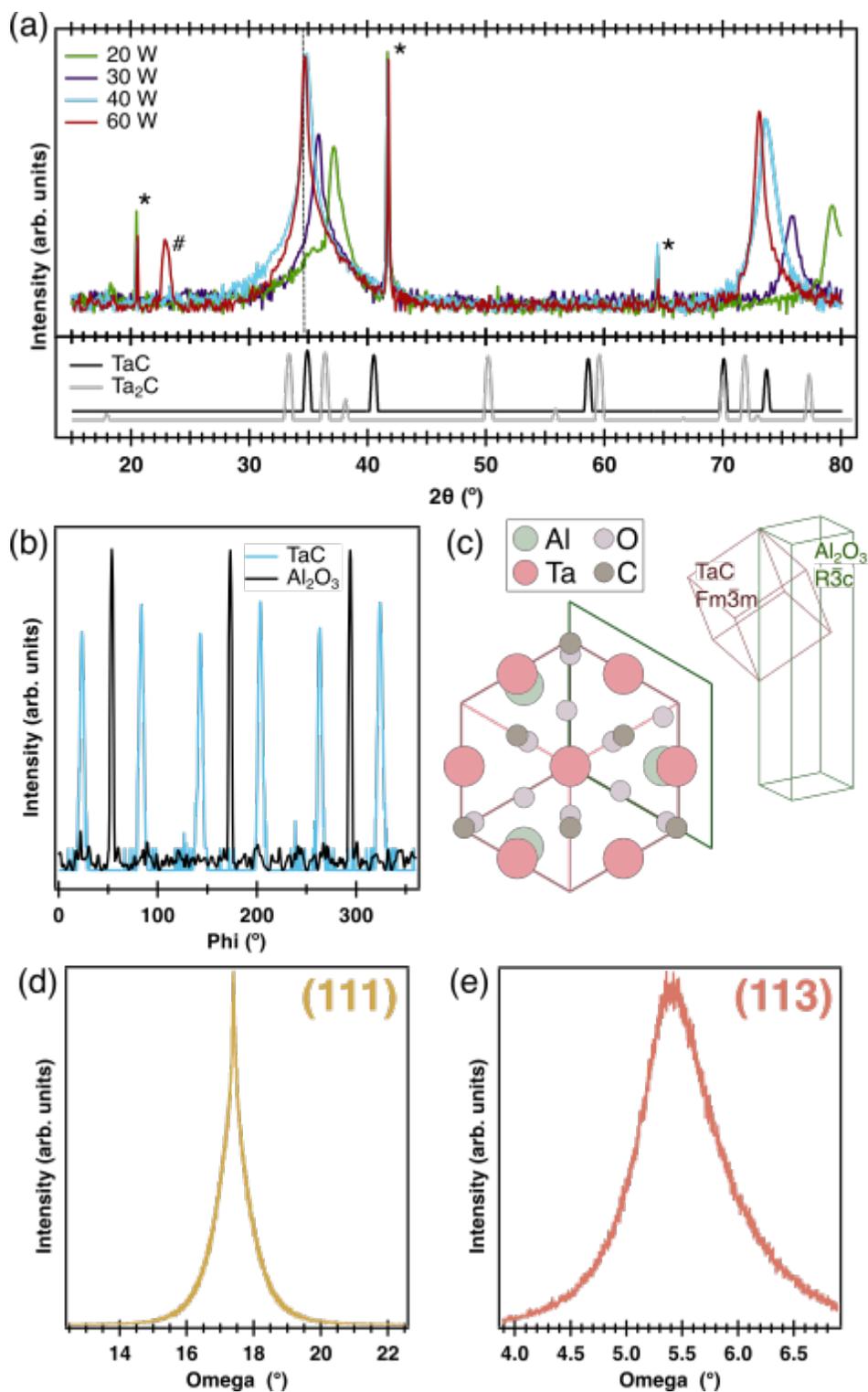

*Figure 2. Crystal structure of sputter deposited TaC films. (a) wide area XRD scan of films grown at 20 W, 30 W, 40 W, and 60 W cathode power along with ICSD references for TaC and Ta$_2$C. (b) rotational alignment of the TaC (002) and sapphire (-102) planes showing coordination between film and substrate. (c) schematic of cubic TaC structure on hexagonal sapphire structure. Dot model shows a top-down view of film coordination. (d) rocking curve of (111) plane in TaC film grown at 40 W. (e) rocking curve of (113) plane of same film.*

*Film composition* | We investigate film composition and microstructure using, transmission electron microscopy (TEM), selective area electron diffraction (SAED), and Rutherford backscattering spectroscopy (RBS). RBS analysis yields a global composition of TaC films containing 4% oxygen, which is the lowest oxygen composition measurable by this system. To corroborate this data, secondary ion mass spectroscopy (SIMS) profiling was performed (see SI) and shows oxygen limited to the top ~10 nm of the film surface, likely resulting from oxidation after growth. Omitting oxygen from the composition of the bulk of the film, we find our films have an average composition of $Ta_{1.0\pm.02}C_{0.95\pm.02}$ across the distance of the wafer, noting that the rocksalt structure of transition metal carbides can tolerate a large degree of substoichiometry.[17,38] From literature we know that films with [C] ≥ [Ta] grown by sputtering result in amorphous carbon segregated out of the film at grain boundaries.[39] RBS also shows the tantalum signal decreases toward the front of the film, likely due to surface oxidation. Electron backscatter diffraction (EBSD) of the same sample (see SI) confirms that all identifiable grains are (111) oriented.

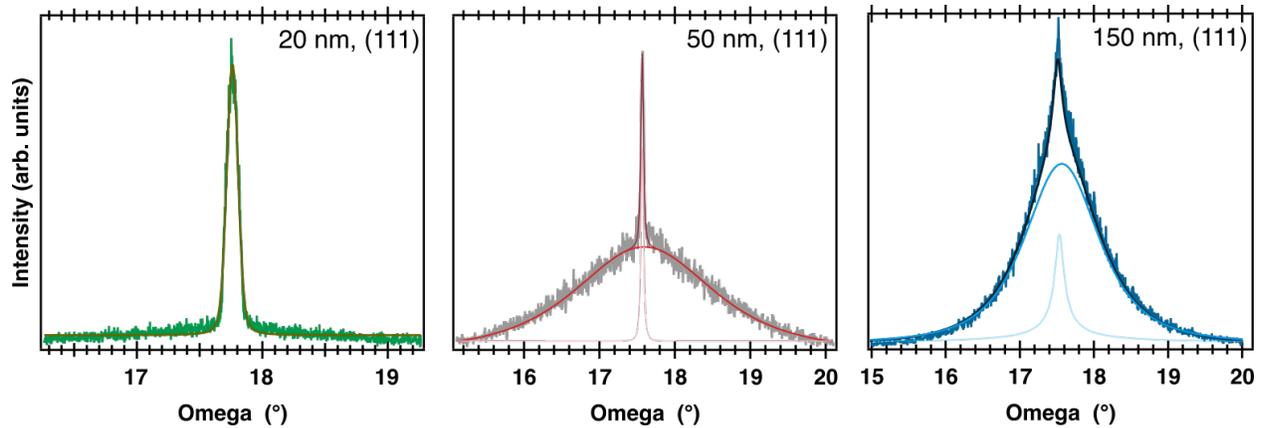

*Figure 3. Effect of film thickness on structure. (111) rocking curve of (a) 20 nm, (b) 50 nm, and (c) 150 nm thick film. Light traces show curve fits breaking each rocking curve into a sharp and broad component.*

*Effect of film thickness* | In high quality optoelectronics, understanding critical thickness is important for minimizing defects and reducing dislocations and film cracking. Literature on epitaxial growth of sputtered films on sapphire shows that grain tilting can be affected by overall film thickness.[36] In AlN-on-sapphire systems, a relaxed layer forms after some critical thickness and is characterized by columnar grains rotated in-plane relative to one another. Further growth results in grains that are still oriented but have increased tilt due to residual stress, manifesting as broader in-plane and out-of-plane XRD peaks; to some extent, these defects can be healed upon annealing at high temperatures. Moreso, film thickness has been shown to control lattice constant and strain of sputtered and annealed AlN films on sapphire.[36] We investigate changes as a function of thickness by growing TaC films with thicknesses between 25 nm and 150 nm for films grown with an indirect incident plasma angle. Thickness and growth rate of the 25 and 50 nm films is verified by x-ray reflectivity (XRR). Modeling yields a growth rate of approximately 70 nm/h at the center of the substrate. This is used to determine growth conditions for the 150 nm film. The modeled density is significantly lower than that expected for bulk TaC, fitting at 11.0 g/cm³ rather than the expected 14.3 g/cm³. **Fig 3(a), (b), and (c)** show rocking curves of the (111) peak for the

25, 50 and 150 nm film respectively. Consistent with literature on AlN virtual substrates,[36] the thinner films have reduced components of tilted and relaxed grains compared to the 150 nm film. In the case of the 30 nm film, only the sharp component could be resolved accurately by peak fitting.

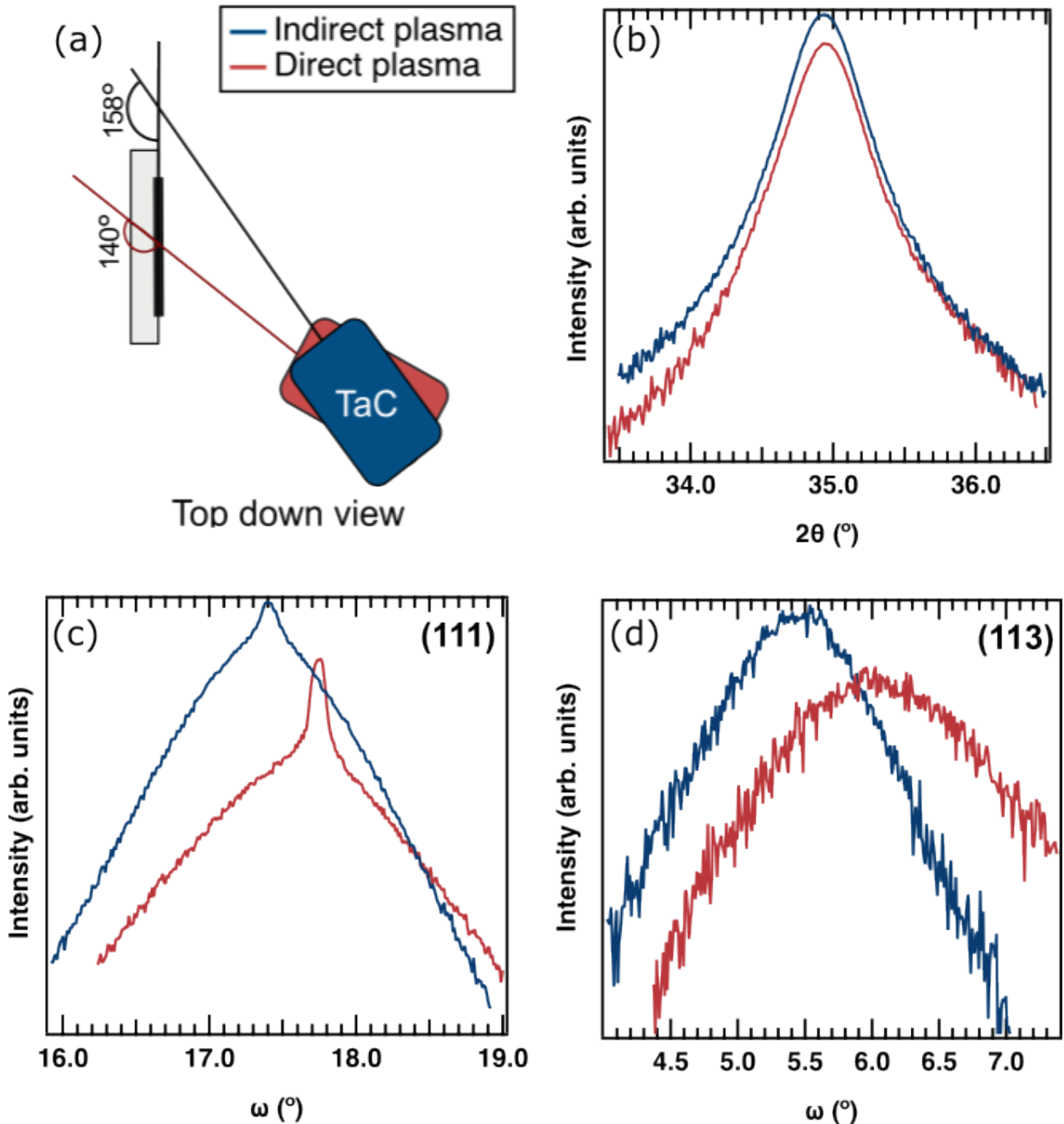

*Figure 4. Effect of incident plasma angle on film structure. (a) schematic of sputter cathode position relative to substrate, denoted by the thick black line. (b) XRD pattern of the (111) TaC peak for a film with direct plasma incidence (red) and indirect plasma incidence (blue). (c) (111) and (d) (113) rocking curves of the films in (b).*

*Exploring effect of incident plasma angle on film structure* | High quality optoelectronics require low density of threading defects and a low propensity for cracking; in many cases, these defects can originate from the substrate.[37] Previous reports show that strain and defect density are sensitive to the angle of incident plasma relative to the growing surface because a plasma pointed directly at the film has higher energy than one with a glancing plasma angle.[40] **Fig 4** shows the structure of films deposited with two different incident plasma angles. The orientation in which the sputter cathode points directly at the substrate is referred to as "direct"; the sample deposited with an "indirect" plasma has the sputter cathode pointing beyond the substrate. A schematic of the growth chamber is shown in **Fig 4(a)**. The film thickness is fixed at 150 nm in both cases, and the films appear to be single phase by symmetric XRD. **Fig 4(b)** shows that for these films the (111) peak position and therefore out-of-plane lattice spacing is nearly identical. In **Fig 4(c)** and **Fig 4(d)** we again utilize rocking curves of the symmetric (111) and asymmetric (113) peaks respectively to 1) confirm the TaC rocksalt phase and 2) identify changes to film lattice constants. **Fig 4(c)** shows that the in-plane structure is similar for both films, but a larger contribution from the sharp peak in the (111) rocking curve for the directly sputtered film suggests fewer tilted grains for the direct case compared to the indirect case, and the a lattice constant of the indirectly sputtered film is 0.04 Å smaller than the directly sputtered film. **Fig 4(d)** shows that the off-axis crystalline quality increases slightly for the film grown with an indirect plasma and the offset angle increases ($\omega$ decreases). The FWHM of the (113) reflection reduces by 33% from 5870 to 3900 arcsec. While the films described in **Fig 4** characterize regions with similar thicknesses, we note that the substrate was not rotated during growth and there is some spatial variation across the film. An example of this is shown in the Supporting Information (SI). Overall, across many depositions, the sputter gun angle has been a useful tool for tuning the crystallinity and phase of the sputtered film. **Table 1** gives a summary of FWHM values and lattice constants for each film.

*Table 1. Strain, peak analysis for films with different thicknesses and incident plasma angles, as discussed in Figs 3 and 4.*

| Sample | Plane | Broad FWHM (arcsec) | Sharp FWHM (arcsec) | d (Å) | $\angle_{113\text{-}331}$ (°) | $\alpha$ (°) | a (Å) |
|---|---|---|---|---|---|---|---|
| 50 nm, direct | (111) | 5490 | 400 | 2.565 | 51.49 | 90.0 | 4.45 |
|  | (113) | 5870 | -- | 1.341 |  |  |  |
|  | (331) | 5360 | -- | 1.022 |  |  |  |
| 150 nm, indirect | (111) | 9340 | 390 | 2.563 | 52.04 | 89.5 | 4.37 |
|  | (113) | 5290 | -- | 1.334 |  |  |  |
|  | (331) | 6290 | -- | 1.016 |  |  |  |
| 50 nm, indirect | (111) | 4330 | 570 | 2.566 | 51.96 | 89.6 | 4.41 |
|  | (113) | 3900 | -- | 1.334 |  |  |  |
|  | (331) | 4490 | -- | 1.017 |  |  |  |
| 25 nm, indirect | (111) | -- | 330 | 2.513 | -- | -- | -- |
|  | (113) | 6900 | -- | 1.321 |  |  |  |
| 225 nm, indirect, 1500 °C annealed | (111) | 3530 | 340 | 2.523 | 51.70 | 89.8 | 4.41 |
|  | (113) | 2240 | -- | 1.333 |  |  |  |
|  | (331) | 2060 | -- | 1.015 |  |  |  |
|  | (111) | 974 | 320 | 2.564 | 51.80 | 89.7 | 4.42 |
| 150 nm, indirect, 1600°C annealed | (113) | 710 | -- | 1.329 | | | |
| | | | | 1.016 | | | |
| | (331) | 790 | -- | | | | |
| TaC reference | (111) | -- | -- | 2.571 | 51.50 | 90 | 4.453 |
| | (113) | -- | -- | 1.343 | | | |
| | (331) | -- | -- | 1.022 | | | |

We next investigate structural changes as a function of film thickness and sputtering condition. Because the TaC plane is epitaxially aligned in the (111) direction, strain due to lattice mismatch and/or thermal expansion mismatch could change the α angle of the crystal away from the cubic 90° condition, more closely resembling a rhombohedral system. To verify if this is the case, we look at the angle between the (113) and (331) planes as these planes can both be accessed without moving or realigning the sample. In a cubic system with α=90, the angle between the planes is 51.50, and the relationship between the interplanar angle, $\angle_{113-331}$, and α is both linear and independent of the a-spacing. To determine the α of our thin films we characterize the (113) and (331) reflections in an asymmetric geometry by XRD without changing the sample alignment or moving the sample stage, move to the asymmetric position of the (331) plane (positive omega offset). For a rhombohedral crystal measured in an asymmetric geometry ($\chi = 0$):

$$\angle_{113-331} = (\theta_{113} - \omega_{113}^-) + (\theta_{331} - \omega_{331}^+)$$

where the (113) reflection is measured with a negative ω-offset (ω⁻) and the (331) reflection is measured with a positive omega offset (ω⁺) without moving or realigning the sample. This more accurately reflects the strain in the crystal than calculating changes to the lattice constant in a cubic geometry. We show the results of these measurements and calculations in **Table 1** and identify a compressive rhombohedral distortion in the in-plane direction, with α angles of about 89.5° for the 50 and 150 nm film deposited using an indirect incident plasma angle. The lattice constant a is then found by fixing α and a least-squares regression of the measured $d_{hkl}$ values.[41] To corroborate these lattice constants, an unbound least squares regression to the measured $d_{hkl}$ values was implemented which confirmed the magnitude and sign of the rhombohedral distortion. For reference, the ideal parameters for stoichiometric TaC are given in **Table 1.** The interplanar angle of the thinnest film could not be determined due to low x-ray intensity counts obscuring the (331) peak position. If we assume that for the 25 nm case α = 89.5°, the a-spacing would be 4.32 Å which is significantly lower than the other measured values. It is likely that the strain is different below the critical thickness for columnar relaxation, and therefore α is not the same as the thicker films; however, this is not measurable with a benchtop XRD due to low counts.

*High temperature annealing for improved film quality* | Literature demonstrates that thin films annealed in a face-to-face configuration result in improved crystallinity, likely due to spatial confinement and a local partial pressure at the film surface impeding reconstruction or evaporation.[42] To improve the crystallinity of our TaC thin films we annealed them at high temperatures in a face-to-face configuration. Annealing was performed in an Ar atmosphere to discourage formation of $Ta_2O_5$, which is highly volatile at our annealing temperatures. **Fig 5(a)** shows a symmetric XRD pattern for TaC films deposited at 40 W and annealed at 1600 °C for 3 hours. The annealed film has a more intense and less broad (111) and (222) peak, suggesting

improved crystal quality with fewer defects and larger grain sizes. Films before and after annealing are aligned to the sapphire substrate and the presence of a high-angle shoulder in the as-deposited film disappears upon annealing. Similar changes to film crystallinity are reported in literature of face-to-face annealed AlN on sapphire, including the loss of the two-component behavior of the rocking curves. **Fig 5(b) and 5(c)** show rocking curves of the (111) and (113) peaks respectively. After annealing, the (111) rocking curve peak increases in intensity and loses most of the 2-layer character, narrowing to 390 arcsec. The (113) rocking curve peak FWHM reduces by >80% from 1920 to 320 arcsec, and both peaks shift to higher omega consistent with the observed decrease in lattice constant. The interplanar spacing, rocking curve FWHM, and structure are tabulated in **Table 1**. In addition to the increased crystalline quality, the $\angle_{113-331}$ angle decreases indicating the rhombohedral distortion has also decreased upon annealing.

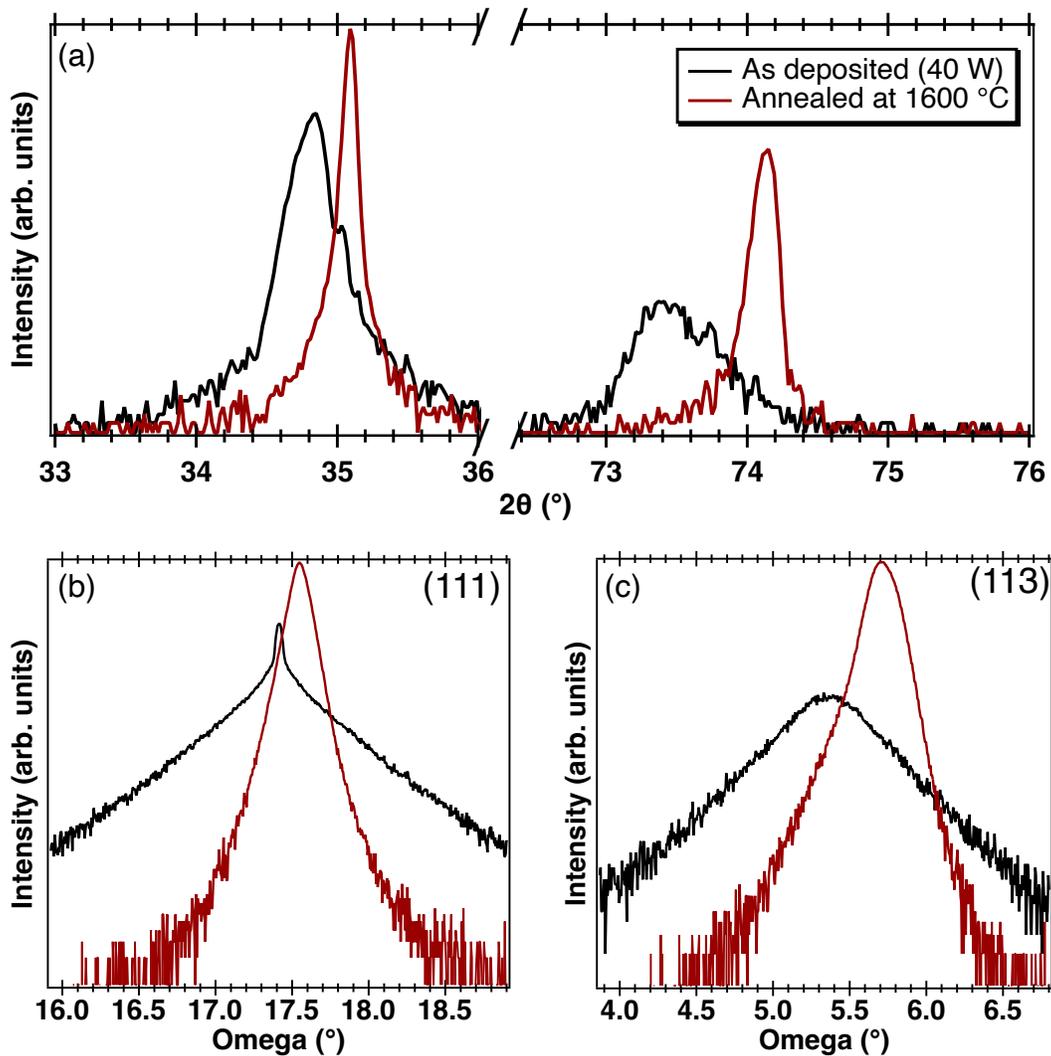

*Figure 5. Annealing of TaC thin films. (a) XRD pattern sample annealed at 1600 °C for 3 hours (red) compared to the as-deposited film. Rocking curve of (b) (111) and (c) (113) peak before (black) and after (red) annealing.*

The surface morphology of the films after annealing is assessed by AFM. **Fig 6(a)** shows the morphology of the TaC film before annealing. The surface shows columnar grains as expected for a sputtered film with a low adatom diffusion length with a roughness of 0.3 nm RMS.[43] **Fig 6(b)** shows the TaC surface after annealing at 1600 °C. Here we see a step-and-terrace surface with approximately 200 nm – 600 nm wide terraces with a terrace slope consistent with the nominally 0.2% unintentional offcut in the sapphire substrate. We primarily observe major steps of 2.5 – 2.6 nm height and minor step heights of about half this value, 1.3 nm, and still observe sub-nm roughness at 0.7 nm RMS. A linescan for the surface in **Fig 6(b)** can be seen in the SI. The crystalline structure of the interface was investigated by HR-TEM in **Fig 6(c)** which shows well-ordered (111) planes in the TaC film epitaxially aligned to the (0001)-direction of the sapphire substrate. The upper and lower insets show a fast Fourier transform obtained from the TaC film and a selective-area electron diffraction (SAED) pattern of the TaC film and sapphire substrate respectively. The SAED pattern illustrates that the TaC film is epitaxially aligned with the sapphire substrate. These results also indicate phase purity of our film with no microscopic $Ta_2C$ inclusions over the sampled region and confirm the epitaxial relationship between film and substrate. TEM also corroborates crystal twinning (see SI) as observed by the XRD ϕ-scan in Fig 2(b). Considering the improvements observed upon annealing, we hypothesize that higher temperatures will further increase the quality of films. The melting temperature of cubic TaC is ~3975 ºC, and thus the annealing temperatures used here reach less than half of the melting point.[44] By contrast, improvements to AlN crystallinity via face to face annealing utilizes temperatures at >75% of the 2200 ºC melting temperature of AlN, which likely contributes to recrystallization and reduction of defects.[36] In all, crystallinity and film roughness improve significantly following annealing at high temperatures. It is likely that either longer anneal times and/or higher temperatures than studied here will result in further crystal quality improvements amenable to growth of high quality $Al_xGa_{1-x}N$.

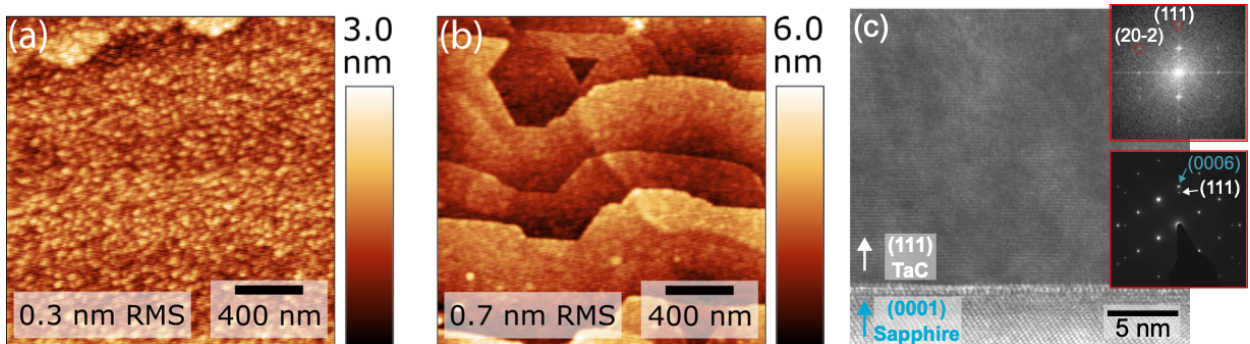

*Figure 6. Morphology and structure of TaC thin films. (a) AFM scan of TaC film as deposited (b) AFM scan of TaC film after annealing. (c) High resolution transmission electron micrograph of annealed film. A fast Fourier transform obtained from the TaC film is in upper inset and the SAED pattern of the film and substrate is in the lower inset*

**Conclusion**

In conclusion, we have demonstrated that epitaxially oriented TaC thin films can be deposited on sapphire substrates by RF sputtering and observe significant improvements to crystalline quality upon annealing at high temperatures. Film structure was investigated to understand the effect of sputter gun power, incident plasma angle, and film thickness on phase purity, lattice parameter, and in- and out-of- plane crystal distortions. Thinner layers deposited by off-axis sputtering have fewer crystalline defects than films deposited by on-axis sputtering, and thicker films show increased relaxation through defect formation, consistent with literature. High temperature annealing significantly improves the FWHM of sputtered films and reduces tilt and twist within grains while maintaining registry with the substrate. A step-terrace surface is observed for annealed films, appropriate for $Al_xGa_{1-x}N$ epitaxy. In summary, the obtained lattice parameter, system tunability, thermal expansion properties, and resistivity of this material make it a promising material class for $Al_xGa_{1-x}N$ heteroepitaxy capable of simultaneously improving vertical device performance and reducing cost.

**Acknowledgements**
The authors would like to acknowledge Eurofins EAG Laboratory (www.eag.com) for collection and analysis of RBS and SIMS data and Patrick Walker for FIB preparation of TEM samples.

This work was authored by the National Renewable Energy Laboratory (NREL), operated by Alliance for Sustainable Energy, LLC, for the U.S. Department of Energy (DOE) under Contract No. DE-AC36-08GO28308. Funding was provided by the Laboratory Directed Research and Development (LDRD) Program at NREL. The views expressed in the article do not necessarily represent the views of the DOE or the U.S. Government.

**Methods**

**Material deposition** Depositions were performed using an RF sputtering system with a resistive substrate heater. The chamber base pressure for all depositions was in the $10^{-7}$ Torr. Substrate temperature was calibrated by bonding a thermocouple to the surface of a sapphire wafer using silver paste, and the maximum growth temperature of 730 °C was achieved at a setpoint of 800 °C. Working deposition pressure was 5 mTorr using a process gas flow of 20 sccm Ar. Sputter cathodes were equipped with a 2" TaC target that is nominally stoichiometric and sits approximately 4" from the substrate surface. Samples were deposited on (0001)-oriented sapphire indium-bonded to a silicon backing wafer; this assembly is bonded using conductive silver paste and clamped to the heater surface. Prior to deposition, wafers were rinsed in acetone, isopropanol, and DI water and then annealed at 1050 ºC for 6 hours with a 5º/min ramp. Sputter gun power was varied between 20 W and 60 W and allowed to cool fully before removing from chamber.

**Annealing** Anneals were performed in a CM furnace under an argon atmosphere. Samples were ramped to temperature at a ramp rate of 5 ºC per minute, annealed for three hours, then cooled at a rate of 5 ºC/min to 200 ºC before being removed from the furnace. The argon flow was approximately 75 sccm flowed through caps on either side of the furnace tube. Caps were cooled using compressed air around a coil machined into the outside of the cap. Substrates were placed on a piece of tantalum foil inside an alumina crucible and annealed face to face with film surfaces touching.

**Characterization** X-ray diffraction and rocking curves was collected using a Rigaku Smartlab using a Cu-K$\alpha$ beam monochromated by a 2-bounce (220) Ge channel cut crystal. The samples were manipulated with a 4-circle Eulerian cradle. X-ray reflectivity scans were performed using the same instrument in parallel beam configuration with Cu-K$\alpha$ radiation. XRR modeling was performed using the Rigaku GlobalFit software and the Motofit package in Igor.[45] Spatial differences across the wafer were collected using a custom mapping routine on the Rigaku Smartlab and analyzed in part using CombIgor, a custom Igor package for processing of high-throughput data.[46]

Rutherford backscattering spectroscopy was performed by EAG Laboratory with a He++ beam at 2.275 MeV and is collected in a two-detector geometry for backscattering angles of approximately 100º and 160º. Fits start with a theoretical model and iteratively adjust until there is good agreement.

Secondary ion mass spectroscopy was performed at EAG Laboratory by means of a CAMECA ims4f magnetic sector instrument with Cs+ primary ions. MCs+ detection has been obtained under Cs+ ions bombardment. During positive secondary detection (MCs+) the beam impinged on the surface with an energy of 5.5 keV and an angle of incidence of 42° with respect to the sample surface normal.

Transmission electron microscopy (TEM) cross-section samples were prepared using a standard lift out technique in a FEI Nova NanoLab 200 dual beam focused ion beam (FIB) workstation. A Pt protective layer was deposited first to protect the sample surface during subsequent FIB sample preparation. 30 kV $Ga^+$ ions were used for most of the preparation followed with a final polish using < 5kV $Ga^+$ ions. Damage remaining from FIB preparation was subsequently removed in a Fischione NanoMill using < 1 kV $Ar^+$ ions with the sample cooled using a liquid nitrogen cold stage. The samples were then examined in a FEI Tecnai F20 UltraTwin field emitting gun (FEG) STEM operated at 200 kV.

Atomic force microscopy (AFM) scans were performed on a Digital Instruments Atomic Force Microscope model in tapping mode. Images were analyzed and prepared for publication using Gwyddion. Background plane subtraction was performed with a 3rd degree polynomial in both horizontal and vertical directions, and no other image manipulation was applied.